\documentclass[12pt,preprint]{aastex}
\def \kms{\ifmmode{~{\rm km\,s}^{-1}}\else{~km~s$^{-1}$}\fi}
\def \vhel{\ifmmode{V_{{\rm hel}}}\else{$V_{{\rm hel}}$}\fi}
\def \vsys{\ifmmode{V_{{\rm sys}}}\else{$V_{{\rm sys}}$}\fi}
\def \vobs{\ifmmode{V_{{\rm obs}}}\else{$V_{{\rm obs}}$}\fi}
\def \degree{\ifmmode{^{\circ}}\else{$^{\circ}$}\fi}
\def \lsun{\ifmmode{{\rm\ L}_\odot}\else{${\rm\ L}_\odot $}\fi}
\def \msun{\ifmmode{{\rm\ M}_\odot}\else{${\rm\ M}_\odot$}\fi}
\def \myr{\ifmmode{{\rm\ M}_\odot{\rm\ yr}^{-1}}\else{${\rm\ M}_\odot$ 
yr$^{-1}$}\fi}
\def \teff{\ifmmode{{\rm{T}}_{\rm eff}}\else{${\rm{T}}_{\rm eff}$}\fi}
\def \mdot{\ifmmode{{\rm\dot{M}}}\else{${\rm\dot{M}}$}\fi}

\newcommand{\ha}{H$\alpha$}

\newcommand{\oil}{[O\,{\sc i}]\ 6300\,\AA}

\newcommand{\nii}{[N\,{\sc ii}]}
\newcommand{\niil}{[N\,{\sc ii}]\ 6584\,\AA}
\newcommand{\niib}{[N\,{\sc ii}]\ 6548\,\AA}
\newcommand{\niiab}{[N\,{\sc ii}]\ 6548,\ 6584\,\AA}
\newcommand{\nev}{[Ne\,{\sc v}]\ 3426\, \AA}
\newcommand{\arab}{[Ar\,{\sc iv}]\ 4711,\ 4740\,\AA}
\def \st{\ifmmode{^{\mathrm{st}}}\else{${^{\mathrm{st}}}$}\fi}
\def \nd{\ifmmode{^{\mathrm{nd}}}\else{${^{\mathrm{nd}}}$}\fi}
\def \rd{\ifmmode{^{\mathrm{rd}}}\else{${^{\mathrm{rd}}}$}\fi}
\def \th{\ifmmode{^{\mathrm{th}}}\else{${^{\mathrm{th}}}$}\fi}
\newcommand{\hnii}{{\rm H}$\alpha+$[N {\sc ii}]}

\lefthead{L{\'o}pez et al.}
\righthead{NGC 6302}

\received{}
\begin{document}

\title{The `Hubble--type' outflows from the high excitation, 
poly--polar planetary nebula NGC 6302.} 
\author{J. Meaburn, J. A. L{\'o}pez and W. Steffen}
\affil{Instituto de Astronomia, UNAM, campus Ensenada,
Ensenada, Mexico.}
\author{ M. F. Graham and A. J. Holloway}
\affil{Jodrell Bank Observatory, University of Manchester, UK}

%\label{firstpage}

\begin{abstract}

Spatially resolved profiles of the \ha\ and \nii\ lines have been
obtained at unprecendented signal--to--noise ratios over the outflowing
lobes of the high--excitation, poly--polar planetary nebula NGC~6302.
A deep image in the light of \niil\ was also obtained of the
extremities of the prominent north--western lobe.
The Manchester Echelle spectrometer combined with the 2.1--m San Pedro
Martir telescope (Mexico) was used for these observations. 

Firstly, an accurate value of the systemic heliocentric radial velocity
of \vsys\ = -29.8 $\pm$ 1 \kms\ has been established. Also, from `velocity
ellipses' across its diameter from previous observations the parallel--sided 
north--western lobe is shown to have a circular section with a tilt
of its axis to the plane of the sky of 12.8 deg. With this starting point
the pv arrays of profiles  have
been very closely simulated, using the SHAPE code, with Hubble-type outflows.
The faint extremities of the north--western outflow are shown to be expanding
at $\geq$~600 \kms.
The prominent lobes of NGC~6302 have then been  generated in an eruptive event
with a dynamical age of
1900 y for the expansion proper-motion distance of 1.04 $\pm$ 0.16 kpc
as measured here by comparing a 1956 image with that taken in 2002.

Kinematical evidence of a  
high--speed `skirt' around the nebular core, 
expanding nearly orthogonally to the lobes, is also 
presented as are the unusual motions at the western extremities of the
NW lobe.

\end{abstract}


\section{Introduction}

NGC~6302 (PN~G349.5+01.0) is a poly-polar planetary nebula (PN), which
was described and drawn as early as 1907 by
Barnard. 
The many estimations of distance to NGC~6302 range from 0.15 to
2.4 kpc though G{\'o}mez et al (1989), from a radio expansion proper motion
measurement  of the bright nebular core give a firm lower limit
to the distance of 0.8 $\pm$ 0.3 kpc (see Sect. 3.3) and give a distance
of 2.2 $\pm$ 1.1 kpc from measurements of the pressure broadening of
radio recombination lines.

The complex morphology of NGC~6302
is clearly evident in the photography of Evans (1959)
and Minkowski \& Johnson (1967); it is approximately bipolar
consisting of two principle lobes. 
A dark lane covers the waist of the nebula and has
prevented the optical detection of the central stellar system. 
This dark lane  has been observed in  H\,{\sc i} absorption
and is bright in a wide range of emissions i.e. 
PAH, free-free radio, Black Body infra-red
and microwave, H\,{\sc i}, H$_{2}$ etc.
(Rodriguez et al 1985; Matsuura, 2005).  
A massive (0.5 -- 3 \msun)
toroid of material must surround and obscure (A$_{V}$ = 5 -- 7 mag) the 
central star with the
line and free-free emission coming from close to the star and the
neutral dusty material causing H\,{\sc i} and optical absorption further
out. Rodriguez (1985) also argue for the existence of an
extended neutral hydrogen halo surrounding NGC~6302.

Feibelman (2001) suggests that the ultra-violet IUE continuum 
spectrum, uncorrected for interstellar extinction, indicates that the central
O {\sc vi}--Type White Dwarf has a G--Type companion. However, although 
Barral et al (1982) detect the same continuum in their own IUE spectrum,
they conclude, after correction for logarithmic interstellar extinction 
coefficients of c between 1.0 and 1.59, that it is scattered radiation
from the hot star. However a symbiotic 
status remains possible  when compared with the similar, though
lower excitation, nebula  Menzel 3 (Mz--3, 
 Bains et al 2004; Schmeja \& Kimeswenger 2001).

A large range of ionization energies
is present in the extended lobes of
the nebula.  Optical
and infrared (IR) line emission has been observed from neutral 
species such as [O\,{\sc i}], through ionized species such as \ha,
[N\,{\sc ii}], up to [O\,{\sc iii}], He\,{\sc ii}, [Ne\,{\sc v}] and even
as high as [Si\,{\sc vii}]
(Meaburn \& Walsh 1980b; Pottasch et al 1985; Ashley \& Hyland 1988;
Feibelman 2001): in fact NGC~6302 is one of the highest
excitation class of planetary nebulae (PNe) known.
As a result, if the nebula is being photoionized by the obscured
central star, the star must be extremely hot ($\geq
10^5$\,K e.g. Pottasch et al 1985; Ashley \& Hyland 1988; 
Cassasus, Roche \& Barlow
2000). A stellar
wind (Meaburn \& Walsh 1980a)  could instead be
shock-exciting the surrounding gas, but Barral et al (1982)
have concluded that radiative ionization is almost certainly
predominant, due to
NGC~6302's close spectral similarity to 
other radiatively-ionized, high-excitation
PNe. The 800 \kms\ broad wings (Meaburn \& Walsh 1980a) 
to the \nev\ line profiles were originally
considered to be direct manifestations of this particle wind: Barral et al
(1982) demonstrate that this is improbable and their origin by 
electron scattering or even as 
instrumental artefacts should be considered. Feibelman (2001) though
identify some very high excitation, coronal--type, 
spectral features that cannot be 
produced by radiative ionization by the central star and propose that
shocks must play a part in their generation. 
The detection of centrosymmetric
polarization of the nebular radiation (King et al 1985) suggests that
the bright core light is being scattered by the dust content
of the outflowing lobes.

The density sensitive \arab\ intensity 
ratio gives an electron density n$_{e}$ $\approx$ 
80,000 cm$^{-3}$ 
immediately adjacent to the dark waist and other density
sensitive ratios, that work in less dense regimes,
show that n$_{e}$ falls to  $\approx$ 1000 cm$^{-3}$ in
the extended lobes of NGC 6302 (Barral et al, 1982). Meaburn \& Walsh (1980b)
though identify a knot in the extemities of the western lobe (Knot 1 
in Fig. 2 b) which has  n$_{e}$ $\approx$ 3000 cm$^{-3}$.

The electron temperature of the  nebula adjacent to the dark waist 
itself has been estimated a
number of times using  diagnostic spectral line intensity ratios.
Oliver \& Aller (1969) found temperatures in the range 15,200\,K
to 22,000\,K at various points over the two principle inner lobes.
Danziger et al (1973) found a temperature of 17,400\,K from 
the IR continuum emission. Aller et al (1981)
observed this brightest portion of the nebula and adopted 16,400\,K
after they found that a number of nebular line diagnostics converge on
this value. These high temperatures (10,000\,K is usually anticipated
for ionised circumstellar gas) are consistent with 20,400 $\pm$ 4000\,K
derived from the relative widths of the \ha\ and \niil\ emission line
profiles
(Barral et al 1982).

Evans (1959) initially suggested that the form of
NGC~6302 suggests dynamic interaction with the ambient medium 
and the kinematics were first
investigated by Minkowski \& Johnson (1967), who found clear
line-splitting in the lobes:  it was immediately obvious that
NGC~6302 is not undergoing spherically-symmetric expansion. 
Minkowski \& Johnson (1967) and
Elliott \& Meaburn (1977) showed that the expansion in the inner
lobes, in particular their neutral \oil\ emitting component, 
is superimposed upon a velocity gradient across NGC~6302. The
kinematics of the nebula
were first thoroughly investigated by Meaburn \& Walsh (1980b)
using longslit spectroscopy.  The distribution of radial velocities
was found to be consistent with the existence of a number of cavities
flowing out from the center of the nebular waist, hence
the poly-polar description of the nebula. The 
dominant `closed' cavity is to the NW  though 
nebular  emission certainly extends in this direction much 
further than suggested in the previous imagery. 
Early indications were found by Meaburn \& Walsh (1980b)
that the  kinematics and morphology
of this extreme region are not easily explicable within a closed lobe
model with simple outflowing walls; 
consequently, new very deep, spatially resolved position-velocity (pv) 
arrays of profiles of the 
\niil\ nebular
emission line over NGC~6302 have been obtained . 
These  new observations have taken advantage of the technical
improvements in the 25 years since those of Meaburn \& Walsh (1980b)
not least a CCD replacing the earlier photon counting detector, the
use of an echelle grating in the spectrometer along with up-to-date
displays of the data.

In these most recent observations particular, though not
exclusive, attention has been
paid to this far tip of the north-western lobe. A deep
\niil\ image has been obtained to highlight the further
extension of this part of the nebula. Morphological and kinematical
modelling has also been performed to elucidate more clearly the
stucture and motions of these nebular lobes. A firm expansion proper
motion distance has been established by comparing a 1956 image
of Evans (1959) with the most recent one obtained in the present work.

\section{Observations and results}

The present observations were made with the Manchester Echelle
Spectrometer (MES-SPM - see Meaburn et al. 1984; 2003) 
combined with the 2.1-m San Pedro Martir telescope on 
2001 17/18 May and  2002, 10/12 May. 
A SITe CCD was the detector with 1024$\times$1024, 24~$\mu$m pixels
although 2$\times$2 binning was employed throughout the observations.

\subsection{Imagery}

MES-SPM has a limited imaging capability with a 
retractable plane mirror isolating
the echelle grating and a clear area (here 4.37$\times$5.32
arcmin$^{2}$) replacing the spectrometer slit.
The image in Fig. 1 is a mosaic of those snapshot images
through the 90 \AA\ bandwidth \hnii\ interference filter
taken immediately before obtaining spatially
resolved longslit spectra of these lines (see Sect. 2.2).
Whereas the image in Figs. 2a \& b is a subset from this larger field
obtained 
on the 2002, 12 May
with this MES--SPM  imaging configuration but 
covering only the western lobe of NGC~6302.
The integration time was 900s.
The coordinates (J2000)
were added using the STARLINK \textsc{gaia} software. The 10\AA\
bandwidth 
interference filter 
that was employed transmits only the \niib\ emission line.

\subsection{Longslit spectroscopy} 

Spatially resolved, longslit line profiles 
were obtained with the MES--SPM
along the lines marked 1--7 in  Fig. 1. The details of
all spectroscopic observations are listed in Table 1. 
In this spectroscopic mode MES--SPM
has no cross--dispersion consequently, for the present 
observations, a filter of 90~\AA\ bandwidth was used to isolate 
the 87$^{\rm th}$ echelle order containing the \ha\ and \niiab\
nebular emission lines.

\begin{table}
 \begin{minipage}{\textwidth}
 \centering
  \begin{tabular*}{\textwidth}{@{\extracolsep{\fill}}cccc}
   \\
   \hline
	Slit Position	
		& Slit Orientation
				& Slit Width ($\mu$m)
					& Exposure Time (s)	\\
   \hline
	1	& EW		& 150	& 300			\\
	2	& EW		& 150	& 1200			\\
	3	& EW		& 150	& 1800			\\
	4	& EW		& 150	& 1800+2400		\\
	5	& NS		& 300	& $2\times1800$		\\
	6	& NS		& 300	& $2\times1800$		\\
	7	& NS		& 300	& $2\times1800$		\\
   \hline						
  \end{tabular*}
  \caption[Slit widths, orientations and exposure times used to obtain
           longslit spectra of NGC~6302]{Slit widths, orientations and
           exposure times used to obtain the new longslit spectra at
           the positions shown in Fig.1.}
  \label{ngc6302_exp_times}
 \end{minipage}
\end{table}

The 512 increments, each  0.624\arcsec\ long, give 
a total projected slit length of 5\farcm32 on the sky. 
`Seeing' was always $\approx 1$\arcsec\ during these observations.
A 150~$\mu$m wide ($\equiv 11$~$\kms$ and 1.9\arcsec) slit was employed
for the brighter nebulosity nearer the nebular core.
For the spectrometry of the faintest but high--speed gas this slit was
changed to one which is   300~$\mu$m wide 
($\equiv 22$~$\kms$ and 3.8\arcsec) to achieve sufficient signal
to noise ratio though at the expense of spectral resolution.  

The spectral data were bias corrected, cleaned, etc. in the usual
way using the STARLINK \textsc{figaro} and \textsc{kappa} software
packages. All spectra were calibrated in heliocentric radial velocity
(\vhel) to $\pm$~1~$\kms$
accuracy against the spectrum of a thorium/argon lamp.

The negative greyscale representations of those parts of
the position--velocity  (pv) arrays 
for slit positions 1 -- 7 are shown in
Figs. 3 -- 10. Only the \niil\ profiles are shown.
These are compared in each case with the snapshot image 
of the sky and slit
taken immediately before the spectroscopic integregation through
the same 90~\AA\ bandwidth filter. The origin of spectral features
are precisely identified by this technique. Sample \ha\ and \niil\
line profiles for positions A and B in Fig. 4 are shown in Figs. 12a \& b;
similarly
\niil\ profiles from positions C -- H in Figs. 5 -- 9 are shown in
Fig. 13.

\section{New kinematical and morphological features}

\subsection{Systemic velocity}
All kinematical features must be discussed with respect to the systemic
heliocentric radial velocity \vsys. The radial velocities of the two
\niil\ line profiles on either side of the central dark
lane and closest to its edges 
for slit position 1 (Fig. 4) afford the opportunity to measure 
\vsys\ with some certainty; they are both single, narrow and reasonably
Gaussian in shape. Gaussian fitting shows the western profile
is centered on \vhel\ = -34.0 \kms\ and the eastern one shown in
Fig. 12a is centered on -25.5 \kms\
with both widths (including instrumental broadening) $\approx$ 31 \kms.
In the assumption that these central velocities are evenly distributed
around \vsys\ then a  value of \vsys\ = -29.8 $\pm$ 1 \kms\ is 
therefore indicated for the whole nebular complex.
The rest wavelength of the \niil\ line is taken as 6583.45 $\pm$ 0.01 \AA\ 
(calibrated against \ha\ at 6562.82 \AA) in these calculations 
(Meaburn et al 1996) and the less accurate  \niil\ rest wavelength
used by Meaburn \& Walsh (1980b) accounts for the $\approx$~ -9 \kms\
difference between \vsys\ given in that paper and the value reported here.

\subsection{The prominent north western lobe}
A starting point for the understanding of the morphology and kinematics
of all of the lobes of NGC~6302, including their
extremities, is the derivation of the detailed form and behaviour of 
the most prominent NW lobe. In the image in Figs. 2 (a \& b) it can be 
seen that for a large part of its length the walls of the  NW lobe 
are parallel.
This is approximately depicted in Fig. 11 a.  The pv arrays in
Figs. 3, 5 \& 6 show clearly both the radial expansion of this 
lobe and the tilt
to more positive radial velocities away from the central star as found
by Meaburn \& Walsh (1980b). `Velocity ellipses' indicative of 
radial expansion with a nearly circular section were
shown clearly by Meaburn \& Walsh (1980b) as depicted schematically
in Fig. 11 b for the A' cut across the NW lobe also marked here in Fig. 2a.
Aspects of this NW lobe of NGC~6302 i.e. its
parallism and velocity ellipses across its width, are found in the lobes
of the comparable object Mz--3 (L{\'o}pez \& Meaburn 1983: 
Meaburn \& Walsh 1985).

The angle $\phi$ = 12.8 deg. can be directly measured off the image 
in Fig. 2b  as sketched in the XY plane in Fig. 11a. 
The separation of the lobe edges along the A' cut was measured
on the image with STARLINK \textsc{gaia} software and simple geometry gives
2$\phi$.
As a circular
section is indicated by the velocity ellipses, and this angle is small,
then to a reasonable approximation  the
structure and kinematics of the lobe in the XZ plane  is shown in Fig. 11c
(the observer is below Fig. 11c).
The measured radial velocity
differences w.r.t. to \vsys, from fig. 4 of Meaburn \& Walsh (1980b)
 are $\delta~V_{1}$ = 0 \kms,  $\delta~V_{2}$ = 
59 \kms\ \& $\delta~V_{3}$ = 117 \kms\ which firstly implies
that the nearside of the lobe along the cut A' is flowing in the plane of the 
sky with expansion velocity V and that the lobe axis is $\phi$ = 12.8 deg.
to the plane of the sky as shown in Fig. 11c.

As V = $\delta~V_{3}$~$\times$~(sin (2 $\times \phi$))$^{-1}$ 
and =  $\delta~V_{2}$ (sin ($\phi$))$^{-1}$ a value
of V = 263 $\pm$ 5 \kms\ is given at the position A' (1.71 arcmin
from the central star) and along the directions shown in Fig. 11c. 
.
The \nii\ image in Figs. 2 a \& b
and the spectacular one by Corradi, obtained with the 3.6--m La Silla
telescope
(http://sci.esa.int/science-e/www/object/index.cfm?fobjectid=34985)
, are also
significant to the dynamical interpretation because these
suggest that the walls of the NW lobe are not uniformally filled
with outflowing gas: long filaments within these walls 
project back to the central star.
Also there are gaps in the pv arrays; most notably where the slit
crosses the western edge of this lobe in Fig. 3.

\subsection{Expansion distance}

After distance (D in kpc) determinations of PNe based on parallax the 
most certain
method is by measurements of expansion proper motions ($\delta\theta$ in
mas y$^{-1}$) particularly when the 
expansion tangential velocities (V$_{t}$ in \kms) are both high and well known.
For the outflow depicted in Fig. 11  V$_{t}$ = V $\times$ cos($\phi$)
then 

\begin{equation}
{\rm D} \times \delta\theta =  {\rm 216.8} \times {\rm V \times cos}(\phi).
\end{equation}

G{\'o}mez et al (1989 - Sect. 1) attempted this measurement of D for the
core of NGC~6302 where the expansion velocity is only 10 \kms. In the
present work
(Fig. 11 \& Sect. 4) the NW lobe velocity is found to be both high (V =
263 \kms\ at A' in Fig. 2a) and directed along
low angles to the plane of the sky.
The expansion proper motion of  $\delta\theta$ = 56 $\times$ D$^{-1}$  
mas y$^{-1}$ is then predicted by the parameters in Sect. 3.2 for
the position A' at 1.71 arcmin from the nebular center in Fig. 2a. 

 Even a simple  visual comparison of the \hnii\ image taken 
Evans (1959) on 1956, 8 August 
of NGC~6302 with the \niil\ image in Fig. 2a reveals significant
angular shifts outwards with time of many of the nebular features
at the extremities of the NW lobe. As the \niiab\ lines combined are
2--3 times as bright as \ha\ in these lobes these images are then reasonably
comparable. The most certain measurement of
this shift is for the knot marked K2 in Fig. 2a for it is
bright and compact and conveniently between two close stellar images. 
 
The position of K2 relative to these stars was made by ruler off an enlargement
of Evan's 1956 published image to an accuracy of 0.5 arcsec. Its position
relative to the same two stars was measured to far higher accuracy using the 
STARLINK \textsc{gaia} software on the data array for the
2002 image in Fig. 2a. A shift outwards is detected along a line
to the nebular center of 3.26 $\pm$ 0.5 arcsec
in the 45.762 y between the two images to give  $\delta\theta$ = 70.4  
$\pm$ 10 mas y$^{-1}$. As the angular distance from the 
core of K2 is 2.24 arcmin
compared with 1.71 arcmin for the position A' and  assuming a Hubble--type
expansion of the lobe (see Sect. 4) then the predicted  
value of $\delta\theta$ = 73 $\times$ D$^{-1}$  
mas y$^{-1}$ for now V = 345 \kms\ 
to give D = 1.04 $\pm$ 0.16 kpc as the distance to 
NGC~6302. Somewhat lower accuracy measurements of knots
and filaments in the vicinity of K2 indicate very similar proper
motion displacements confirming that those of K2 are not anomalous.
Obviously, if the original 1956 plate can be scanned and the results
compared in detail with the data array for Fig. 2a both a more 
accurate distance can be established and the Hubble-type nature of the
outflow tested independently of the kinematical modelling in Sect. 4.

\subsection{The lobe extremities}
The complexities of the extremities
of the NW lobe are revealed in the pv arrays of \niil\ profiles
in Figs. 7 -- 10 as well as in the deep image in Fig. 2b. Separate
expansions across this region are particulary apparent in Fig. 7
with values of \vhel\ reaching $\approx$ 175 \kms\ in Figs. 8 -- 10.
which is $\approx$~205 \kms\ w.r.t. \vsys (and
see the \niil\ profiles in Fig. 13 for position G \& H). This behaviour is
obviously more complex than that depicted for the principal
NW lobe in Fig. 11. Radial expansion of a single coherent feature no longer
seems present, because separate `sheets' of material with different radial
velocities can be seen in the pv arrays. This extremity is westward 
of the group of knots of which Knot 1 (arrowed in Fig. 2a)
is a member. The more eastern knot has a  radial velocity close to \vsys\
(Fig. 6 and \niil\ profiles for
positions E  in Fig. 13) and it
appears to be the region where the prominent NW lobe
changes its character. Knot 1 (position F in Fig. 13 and Fig. 6) though
has a significantly more positive velocity.

\subsection{High--speed `skirt'} 

High--speed velocity components in the \niil\ profiles 
are revealed in Fig. 5 and in the profile
marked C in Fig. 13. Positive velocity
wings are also present in the profiles in Fig. 12a
from position A in Fig. 4. 
These continuous high speed features are most likely 
originating within diffuse material
near the nebular core. The broad velocity 
feature from position C extends to \vhel\ = 120 \kms\ where 
the slit crosses a diffuse `ring'  seen in the 3.6--m  image of Corradi.
This behaviour is very similar to that observed for Mz--3 (fig. 3 of
Meaburn \& Walsh 1985 and Santander et al, 2004 ) 
where  a `skirt'
of material expanding around the nebular  core but nearly 
{\it orthogonally} to
the axis of the main bi--polar lobes is shown to be present. 
A similar situation could prevail
in NGC~6302. There is also a similar morphological structure
in the ejected nebulosity surrounding the LBV star Eta Carinae
(Smith, 2002)
which may suggest that this is a fundemental property of such
outflows.

\section{Modelling the lobes with the SHAPE code}

With the salient parameters of the NW lobe depicted 
in Fig. 11 as a starting point
the SHAPE code 
(Steffen, Holloway \& Pedlar, 1996; 
Steffen, L\'opez \& Riesgo, 2005)
has been used to reproduce as  realistically as possible
the kinematics and morphology of the broader structure of NGC~6302.
In these simulations of the pv arrays
along slits 1--4 a Hubble-type outflow is assumed where V in Fig. 11
is always proportional to the distance from the central stellar
system and pointing along a vector away from it and the
measured (Sect. 3.2) value of $\phi$ = 12.8 deg.
Note that this is a purely morphological and kinematical
model meant to establish the current structure and velocity `field'
in the object. No hydrodynamics or radiation transport has been
calculated.

%The simulations of the pv arrays (Figs. 3, 5 \& 6) across the
%NW lobe  
%are shown in Figs for slit positions 1-3 (Fig.1). 

%In Fig. 14 all of the observed pv arrays are superimposed (slits 1--4
%observations) and shown as
%a function of distance from the central stellar system and remarkably
%even the extremities of the NW lobe (covered by the western ends of
%slits 3 \& 4 and by slits 5--7) show a linear change in radial velocity
%versus distance from star hence a Hubble--type flow is confirmed for
%the whole system. The simulation with this included in Fig. 14 (slits 1--4
%model) 
%convincingly confirms this behaviour.

The modelling with the SHAPE code is performed in two steps. 
First, the structure and 
kinematics is set up in 
the commercially available software {\em 3DStudioMax}, Version 7, 
({\em 3DStudioMax} is a registered trademark of {\em Autodesk, Inc.}).
One might chose any other similar 3D-modeling software. {\em 3DStudioMax}
is the one we consider most suitable for our purpose.
The model data are then rendered as an image and a long-slit spectrum
(pv array)
in a renderer written for the purpose. 
%The general modelling strategy for NGC~6302 was the following. 
%The object structure
%was first approximated by surfaces, which should be as few and as simple
%as possible consistent with the image and spectra. Brightness maps were
%then applied to the surfaces using the material editor feature of 
%{\em 3D Studio Max} which allows virtually any brightness distribution 
%to be applied. This brightness distribution was used to control the
%distribution of particles which emerge from the surface and are allowed
%to travel a small distance such that the surface has a finite thickness.
%We have used
%the {\em ParticleFlow} particle system, which allows the local number
%density of particles to be proportional to the brightness distribution
%mapped on the surface. 
%The particles were assigned a velocity, which in  general
%may be determined by a variety of tools. 
For our
model of NGC~6302 we assumed a velocity proportional to distance
and directed radially outwards, i.e. a Hubble-type velocity law. 

%The position and velocities of the particles were then exported from 
%{\em 3D Studio Max} and rendered 
%as volumetric spheres with radii smaller than the `seeing'
%by  {\em SHAPE}. 
%The images and pv-diagram are displayed within a graphical interface in 
%{\em 3D Studio Max} designed for the purpose. 
%From this interface we do also adjust further input
%parameters for {\em SHAPE}, such as object orientation, colors, 
%particle size and 
%relative emissivity as well as observing parameters like slit position 
%and width,
%seeing and velocity resolution. For direct comparison with observed images,
%the synthetic images and pv-diagrams are then convolved with Gaussian 
%profiles with widths corresponding to the observing conditions.
%For a further detailed description of the software and modelling procedure
%see the forthcoming paper by Steffen \& L\'opez (in prep.).

The structure of the lobes of NGC~6302 
has been synthesised by outlining them with splines
on the direct images in Figs. 1 \& 2
and rotating them around the symmetry axis of each structure, thereby 
producing 
rotationally symmetric lobes. Since the symmetry axes are close to the 
plane of the sky, the deviations of the true outline from the projected one, 
are expected to be very small.
The alignments of the 2 narrow and 2 wide lobes have then been 
adjusted such that they coincide best with the direct image
and the pv arrays.
The inclination of the whole structure with respect to the plane of the
sky has been adjusted according to the axis of the prominent NW lobe, 
for which we have used the value of 12.8 degrees away from  the observer
as determined in Sect. 3.2. 

Two additional structures have been included, the knotty area marked as
K1 in Fig. 2 and the knotty structures in the middle of the southern
lobe. They have been produced each by one spheroidal surface which has
been deformed applying the noise modifier feature of {\em 3D Studio Max}
in order to break the spherical symmetry and obtain a more random 
distribution throughout the volume. The southern structure is located 
entirely within the opening angle of the southern wide lobe.
%The final knotty distribution of particles has been obtained by applying
%smooth, outwardly decreasing brightness gradients combined with a random 
%noise distribution. 
%The scale size of the noise was chosen to be similar to the observed sizes 
%of knots and filaments. 
No attempt was made at this time to reproduce all 
of the
individual small scale structure, although a number of synthetic  
knots by chance 
do coincide with observed knots and serve as a reference. This means that
the `weather conditions' in NGC~6302 and in the model are different.

The predicted pv arrays for slits 1--4, after this process, are shown 
separately in Fig.
14 for comparison with the observed arrays in Figs. 3--7. The 
kinematical correspondence between the predicted and observed
pv arrays is remarkably accurate. Also shown in Fig. 14
is the superposition of the observed pv arrays for slits 1--4 for
comparison with the modelled counterpart. The linear tilt of the axis 
of the outflow in these pv arrays 
for the prominent northwest lobe is clear. 
There is only a small
change in angle to the west of the knots at its extremity 
(of which Knot 1 in Fig. 2
is a member).
The simulations in Fig. 14 therefore 
confirm conclusively 
that the basic outflow
is Hubble--type but suggest that this flow is modified somewhat
after passage through knotty material to the extreme west of this lobe.
This conformity with a Hubble--type flow pattern
reinforces confidence that this present model is a true
representation of NGC~6302 for such flows are found in a wide
range of circumstellar phenomena.

SHAPE modelling permits for the first time
different predicted views (Fig. 15a--d) of the lobes of NGC~6302 along
various axes; similar objects at less advantageous orientations (`nature'
is random in this respect) can then be recognised. These ilustrations
should then help when determining the structures of other poly--polar
PNe from the multitude of complex shapes found when these are imaged.

The first two images 
are taken along the line of sight (Fig. 15a \& b)
and are the normal view of the object, whereas those in Fig. 15c \& d
are views from within the plane of the sky along the north-south and east-west
axes respectively. 
The color coding for Fig. 15a corresponds to the red-blue Doppler shifting 
whereas for the rest the color coding is arbitrary
to allow the separate regions of the object to be  distinguished.
%A total of $10^5$ particles have been used for the model.
%The emission from each particle is integrated along the line of sight,
%assuming that the nebula is optically thin.

\section{Wind driven versus ballistic outflows}

There are two leading possibilities for the creation of the bi--polar
lobes of NGC 6302. In the first by Cant{\'o} (1978), Meaburn \& Walsh (1980b) 
and Barall
\& Cant{\'o} (1981), an isotropic particle wind from the
central star embedded in a dense disk first forms 
bi--polar, pressure--driven, 
cavities perpendicular to this disk. These cavities are
delineated by standing shocks across which the wind refracts  
to drive outflows parallel to the cavity walls.
Alternatively, the bi--polar lobes are simply the ballistic consequences 
of an explosive event in the central stellar system.

The wind model has many attractions, not least that it predicts the dense
knots (K1 etc. Fig. 2a) found at the `acute tip' (Barral \& Cant{\'o} 1981) 
of the
NW lobe of NGC~6302. Within this model the outflowing cavities in the 
extremities of the NW lobe ( Figs.7--10) are predicted as the wind--
driven, out--flowing, cavity walls first collide at this tip 
then expand outwards again
at higher speed.

However, several objections to the wind--driven model have since arisen both
here and elsewhere. Most significantly, Feibelman (2001) finds only 
direct evidence of a very weak
wind in the nebular core with a low terminal velocity of -520 \kms.
Meaburn \& Walsh (1980b) show that for a distance to NGC~6302 
of $\geq$ 150 pc such a weak  wind would be
insufficiently
energetic to form the cavities and then drive the bi--polar outflows (the 
the distance of NGC~6302 given in Sect. 3.3 is 1.04 kpc). The observed 
fragmentation
of the lobe walls would surely also indicate that wind formation 
of pressure driven cavities has not
occurred. 
Furthermore,
the velocity vectors as depicted in Fig. 11 and used throughout the
convincing modelling illustrated 
in Fig.14 point back to the central star even where
the lobe walls appear parallel (e.g. A' in Fig. 2). The `velocity ellipses'
in the pv arrays across the lobe diameters
(Meaburn \& Walsh 1980b) are a direct consequence of this behaviour 
and could not be produced
by flows along the parallel walls of the wind--driven cavity
defined by standing shocks. Elongated wind--driven 
cavities still in a state of pressure--driven 
expansion, combined with wind--driven flows parallel
to their walls, need exploring dynamically to see if
the present observed kinematics can be generated i.e.
`velocity ellipses' across the lobe diameters combined with
Hubble--type outflows.  

However, the `Hubble--type'
velocity field predicted by the modelling in Fig. 14 combined with
these vectors pointing back to the central star favours the formation
of the NW lobe by a single eruptive event. 
A dynamical age for this event of D (kpc) $\times$ 1,800 y 
is determined for V = 263 \kms\ at A' in Fig. 2a \& 11. 
With the value of D = 1.04 kpc measured in Sect. 3.3 then this age
becomes $\approx$ 1,900 y.
This suggestion
though leaves the creation of the complex motions at the extremities
of the NW lobe (west of K1 in Fig. 2a) 
not easily explained other than that they could perhaps be 
a consequence of the fastest parts of the same eruptive event
(Sect.4) overunning pre--existing dense knots. A simple
consequence of the Hubble--type flows revealed in Fig. 14 is that
the outflow velocity will reach V $\approx$~460 \kms\ (Fig. 11) at the
furthest extent of the NW lobe 3.0 arcmin from the central star.
Beyond this, at 3.9 arcmin from the central star along the same axis,
the Hubble-type law prevails but with a steeper slope in which
case V $\geq$ 600 \kms\ is predicted for the faintest
extremities of this NW outflow from NGC~6302.
This is similar to the high maximum speeds ($\geq$ 500 \kms) 
found for the bi--polar lobes
of He2--111 (Meaburn \& Walsh 1989) and for the ejected knots
from MyCn 18 (Bryce et al. 1997).

Incidentally, the bi--polar lobes of the PN Mz--3 
(Meaburn \& Walsh 1985)
have exactly the same characteristics as those of NGC~6302, i.e.
velocity ellipses are found across their lobe diameters and the lobe outflows
are Hubble--type (see fig. 4 of Meaburn \& Walsh 1985 and 
Santander et al. 2004). In the case of Mz--3
any wind could not impinge on the lobe walls for there are dense, inner,
slowly expanding
shells shielding the outer extensive lobes. An ejection event within
the central stellar system shown to possibily consist of two
stars by Smith (2003) with a symbiotic nature suggested
by Bains et al (2004), is similarly 
favoured for the creation of the Mz--3 lobes.
 
The nature of the central source of NGC 6302 remains
unknown. The extremely high
electron temperature in the nebula points towards a very massive progenitor,
though Cassasus et al. (2000) estimate a progenitor mass between 4 and 5 solar
masses. Recently Matsuura et al. (2005) detected an
infrared source at the location of the central star of NGC~6302 
but were unable to
obtain additional information on the source's nature. Although there are
no direct indications of  the central source being a binary
system, the complex morphology of this nebula with several systems of
bipolar axes tilted with respect to each other, such as in the case of NGC
2440 (L{\'o}pez et al. 1998), 
and its eruptive nature revealed here
certainly suggests this possibility which should be
explored further in the infrared regime with adaptive optics techniques.

\section{Conclusions} 

The prominent NW lobe of NGC~6302 has a circular section whose
walls are shown to follow very precisely a Hubble--type outflow.

At 1.71 arcmin from the central stellar system the flow velocity
of this NW lobe
is measured as 263 \kms\ with a lobe tilt to the sky of 12.8 deg.

A nebulous knot in the NW lobe, at 2.24 arcmin from the center, 
has a proper motion outwards of 70.4 $\pm$ 11 mas y$^{-1}$.
The consequent expansion proper motion distance to NGC~6302 is
then 1.04 $\pm$ 0.16 kpc.

An eruptive event 1,900 y ago created the prominent NW lobe
and possibly many of the other lobe structures.

High positive velocity wings to the \niil\ line
profiles to the north of the dark waist of NGC~6302 suggest the
presence of an orthogonal `skirt' as found in Mz--3 (and Eta Carinae).

The western extremities of the NW lobe exhibit a small change in the
slope of the Hubble--type flow suggesting that a collision
with pre--existing globules of gas has occurred which modifies the outflow.

The outflow velocity at the extreme position of the NW lobe reaches
$\geq$~600 \kms.

\clearpage

\begin{figure*}
%\epsfclipon
\centering
%\mbox{\epsfxsize=6in\epsfbox[112 27 389 209]{f1.eps}}
\plotone{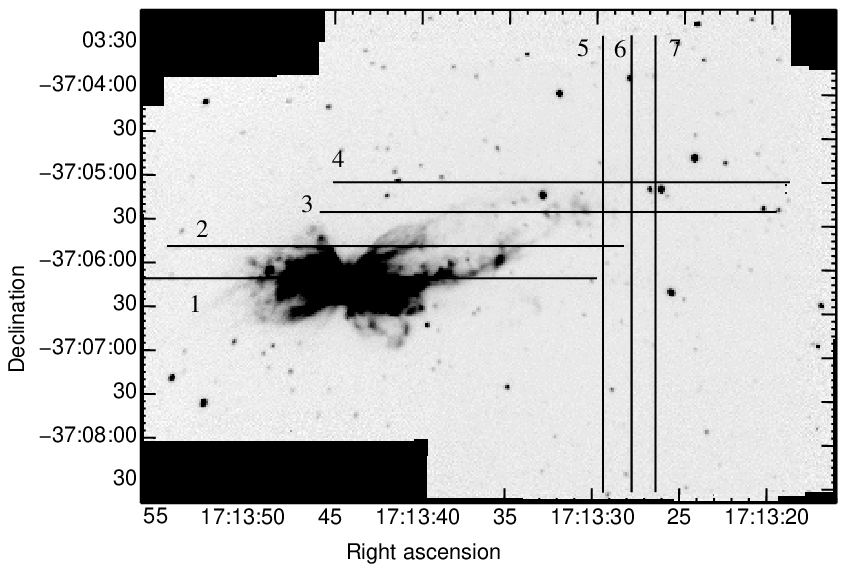}
\caption{The slit positions 1 -- 7 are shown against
a negative image of NGC 6302 in the light of \hnii. (coords
epoch 2000)}.
\end{figure*}

\begin{figure*}
\plotone{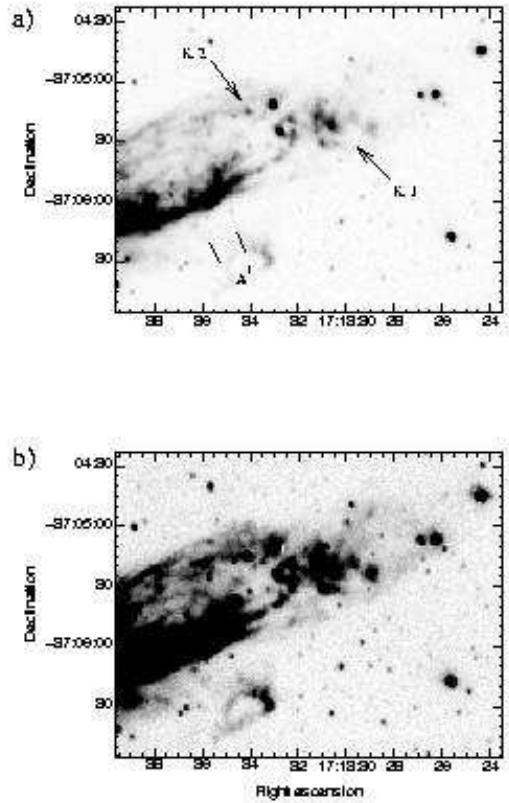}
%\epsfclipon
%\mbox{\epsfxsize=4in\epsfbox[125 27 388 401]{f2.eps}}
%\vspace{-2in}
\caption{An image of the western lobe of NGC 6302 (2002,12 May) through
a narrow filter centered on \niil: a) is a light presentation
to reveal the brighter features. The high--density
Knot 1 from Meaburn \& Walsh (1980b) is marked K1 and 
the band, A', which gives the velocity ellipse in Fig. 11b is indicated.
The outward proper motion of knot 2 (marked K2) has been measured. 
b) A  deep presentation
of the same image to reveal the faintest structure.}
\end{figure*}

\begin{figure*}
\epsscale{1.0}
\plotone{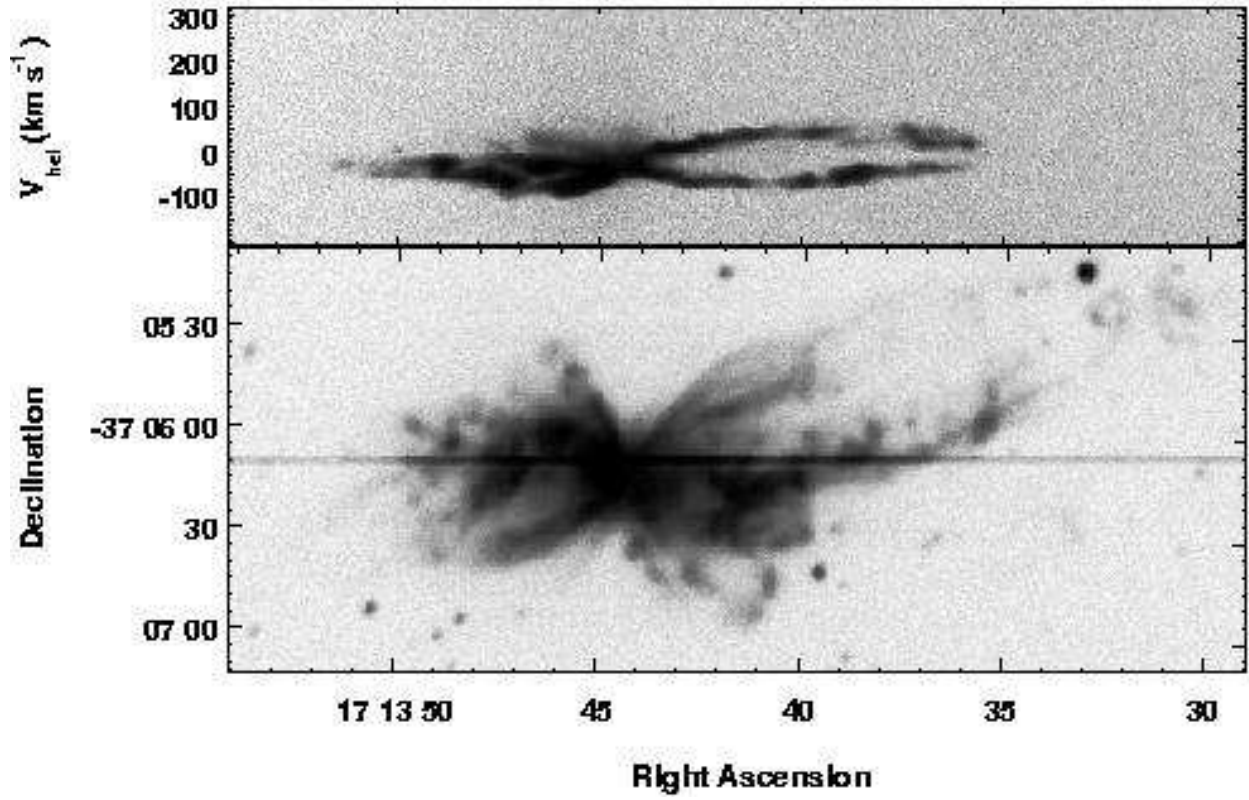}
%\epsfclipon
%\mbox{\epsfxsize=3in\epsfbox[0 0 442 285]{f3.eps}}
\caption{The pv array of \niil\ profiles for the EW slit position 
1 is shown in the top panel.
This should be compared in detail with the image of the slit across
the \hnii\ image of NGC 6302 in the bottom panel.
  }
\end{figure*}

\begin{figure*}
\epsscale{1.0}
\plotone{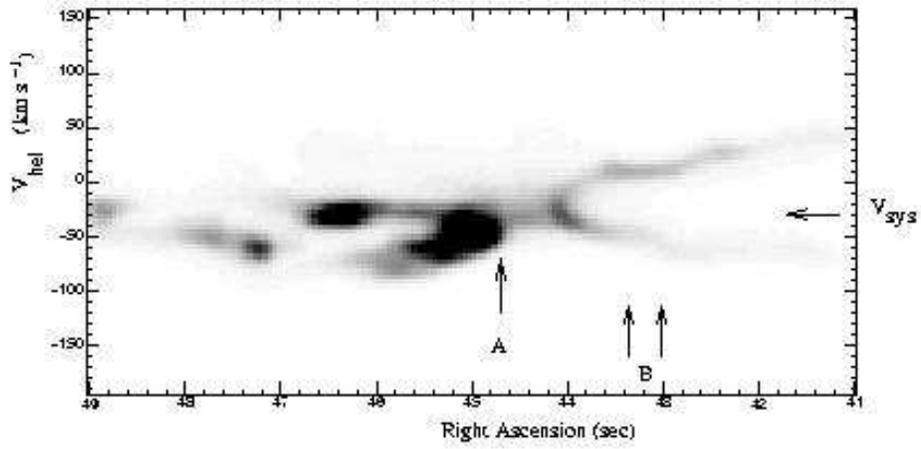}
%\epsfclipon
%\mbox{\epsfxsize=3in\epsfbox[-48 205 644 577]{f4.eps}}
\caption{A light presentation of a subset of the pv array
of \niil\ profiles
for slit postion 1 in Fig. 3 that crosses the nebular core.
\vsys\ is arrowed. The positions where the line profiles
in Figs. 12a \& b were obtained are marked A and B.
}
\end{figure*}

\begin{figure*}
\epsscale{1.0}
\plotone{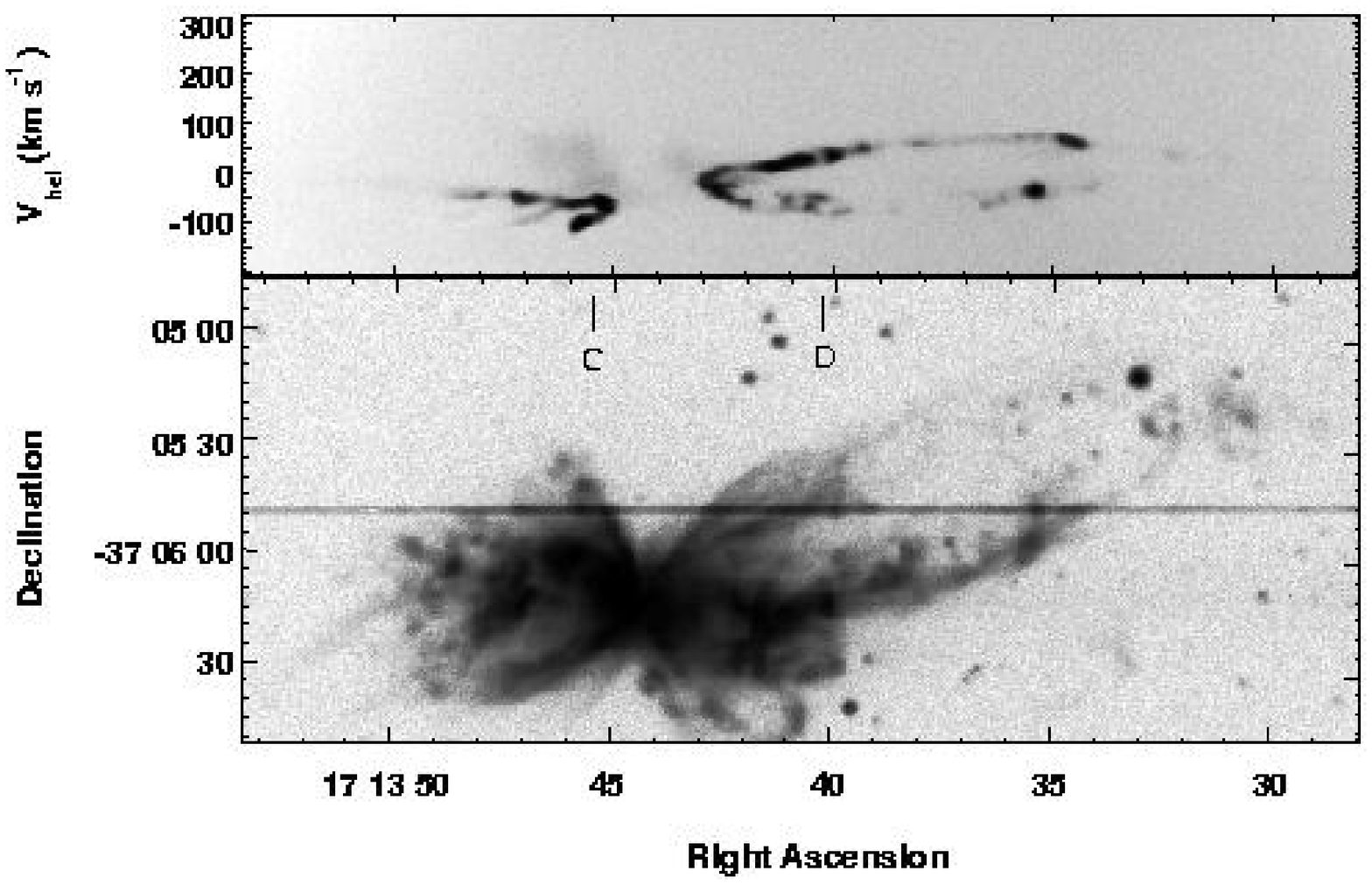}
%\epsfclipon
%\mbox{\epsfxsize=3in\epsfbox[0 0 442 285]{f5.eps}}
\caption{As for Fig. 3 but for slit position 2.The profiles from C \& D
are shown in Fig. 13.}
\end{figure*}

\begin{figure*}
\epsscale{1.0}
\plotone{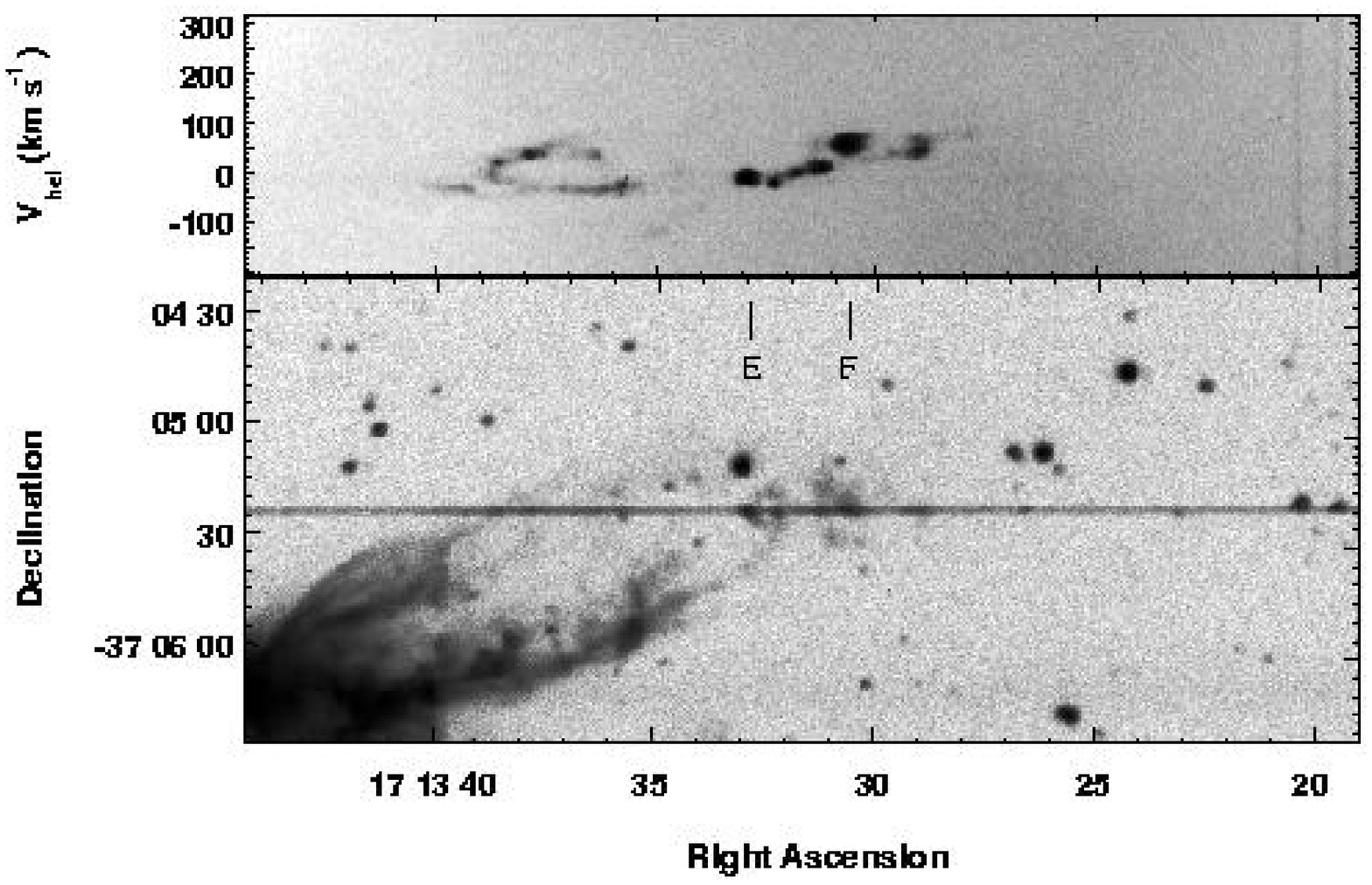}
%\epsfclipon
%\mbox{\epsfxsize=3in\epsfbox[0 0 442 285]{f6.eps}}
\caption{As for Fig. 3 but for slit position 3.The profiles
from E \& F are shown in Fig. 13.}
\end{figure*}

\begin{figure*}
\epsscale{1.0}
\plotone{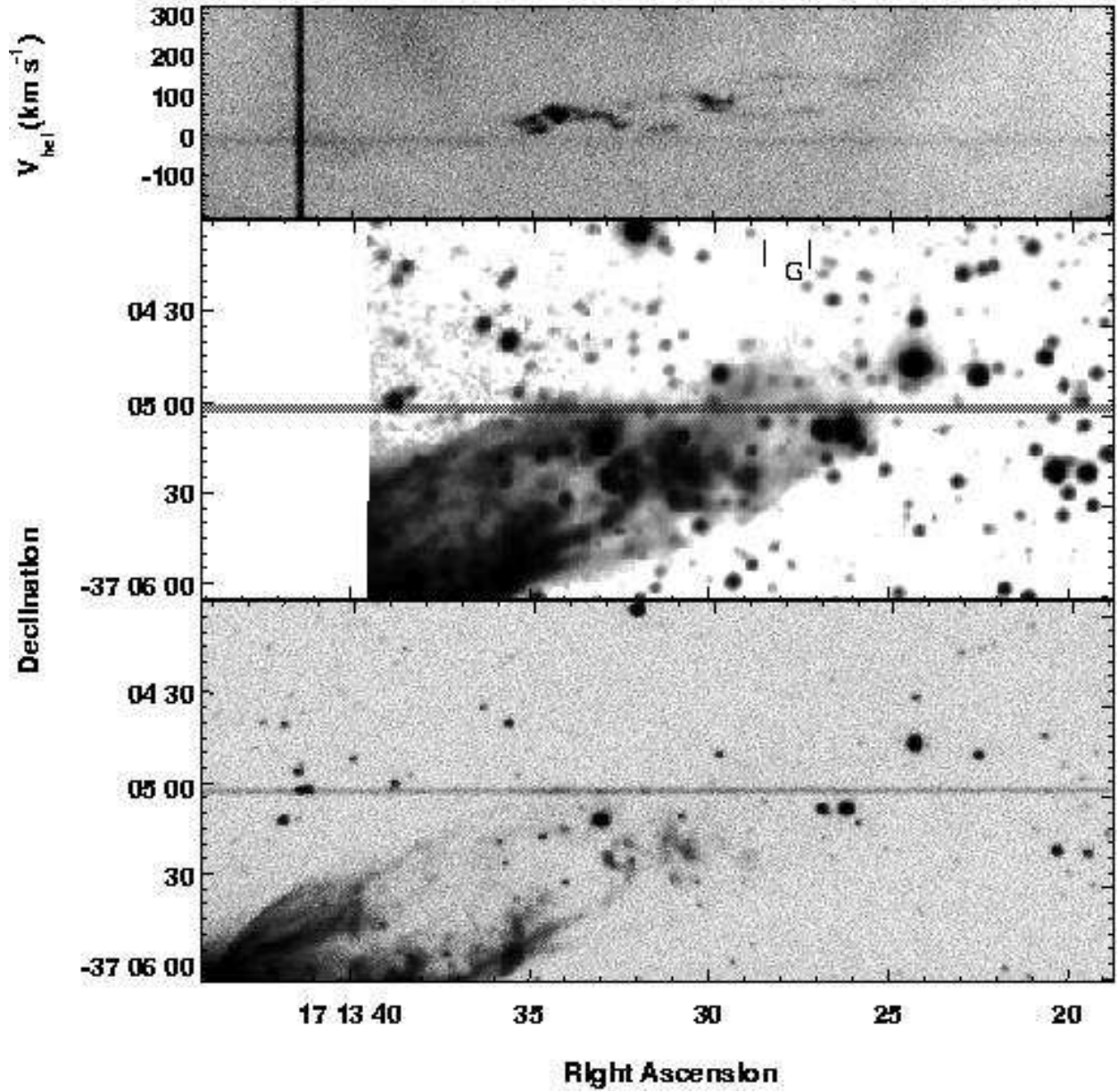}
%\epsfclipon
%\mbox{\epsfxsize=3in\epsfbox[0 0 442 436]{f7.eps}}
\caption{As for Fig. 3 but for slit position 4. The
profile from the band G is shown in Fig. 13. The central panel
is a deep image taken at the time of the spectral observation.}
\end{figure*}

\begin{figure*}
\epsscale{1.0}
\plotone{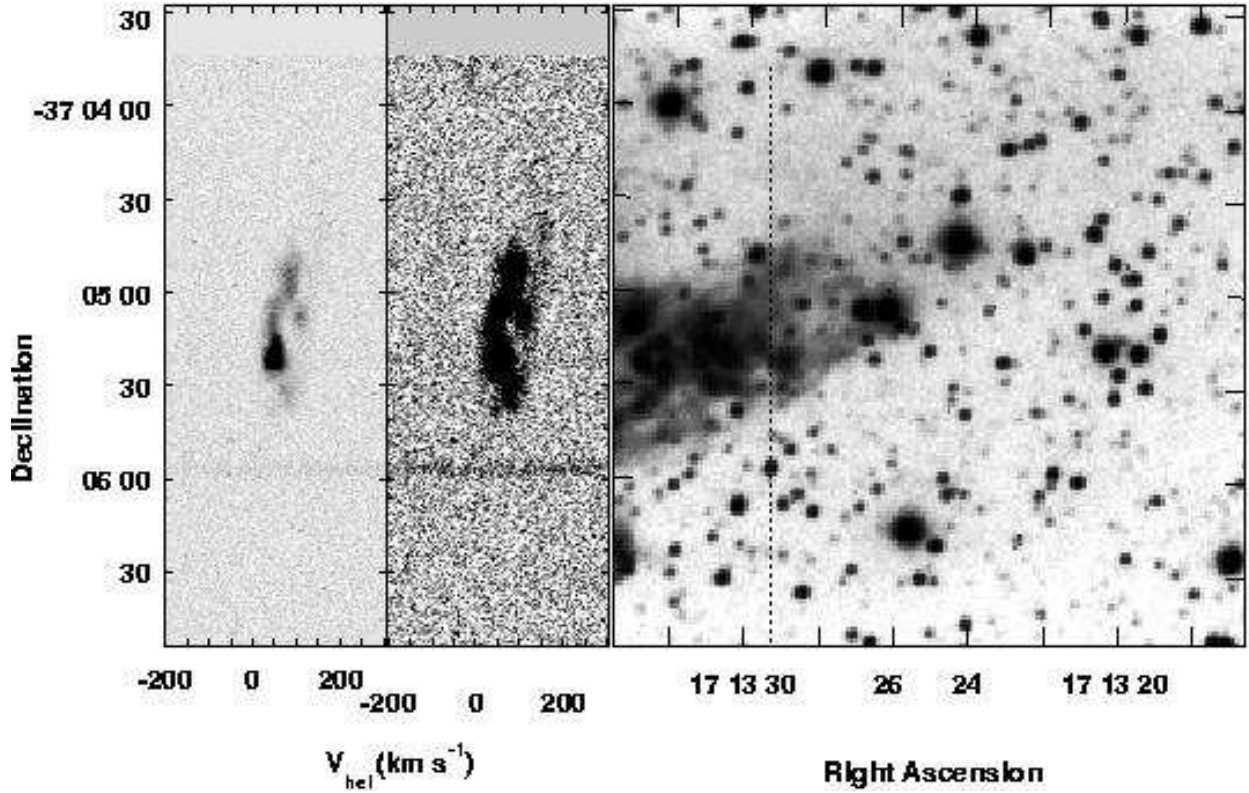}
%\epsfclipon
%\mbox{\epsfxsize=3in\epsfbox[0 0 482 308]{f8.eps}}
\caption{A light and deep presentaion of the \niil\ pv array
of profiles along the NS slit position 5 are shown in the two
left handside panels. These should be compared to
slit position shown as a  dashed line against the \hnii\ image
of NGC 6302 in the right handside panel.}
\end{figure*}

\begin{figure*}
\epsscale{1.0}
\plotone{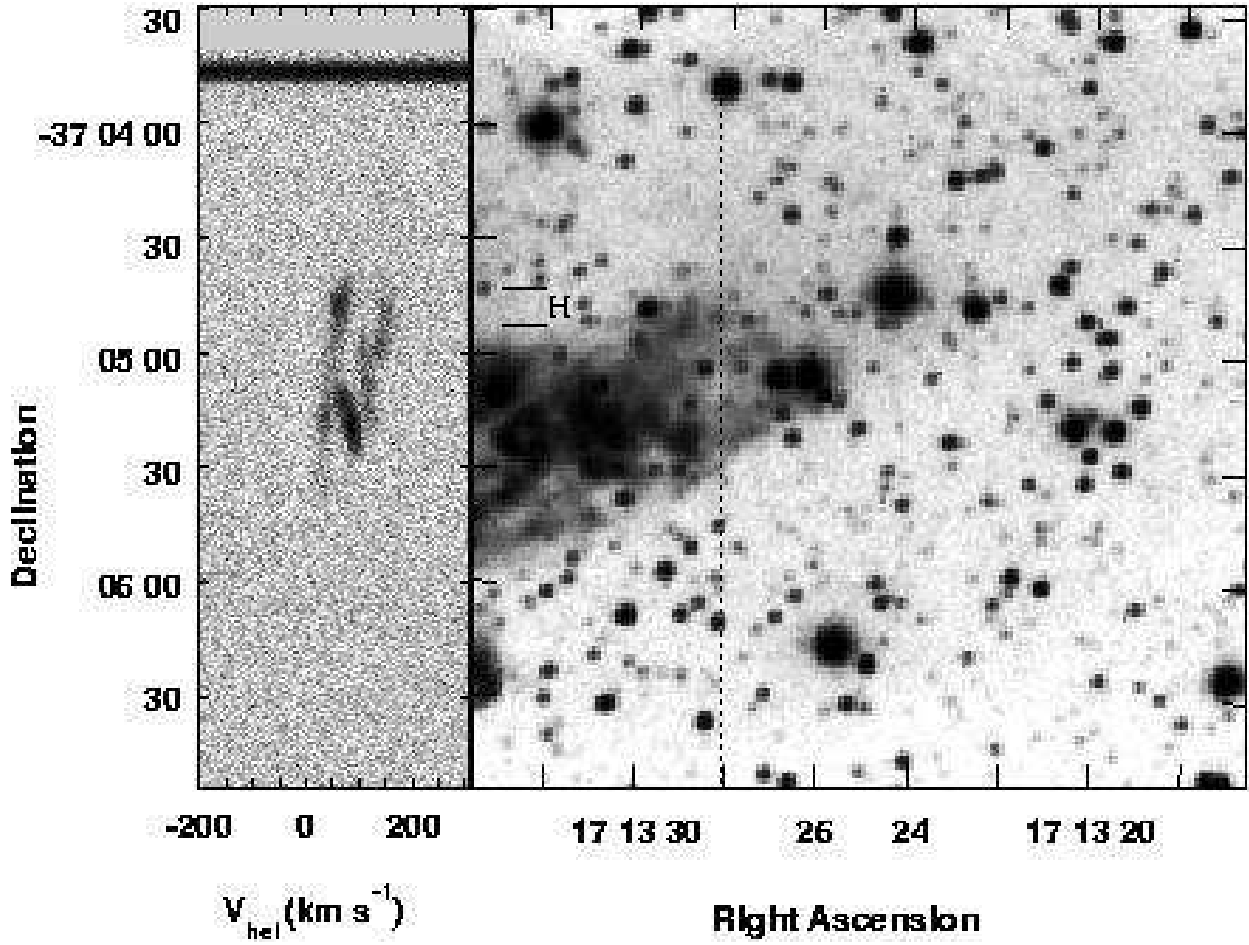}
%\epsfclipon
%\mbox{\epsfxsize=3in\epsfbox[0 0 395 299]{f9.eps}}
\caption{As for Fig. 8 but for slit position 6. Only one presentation of
the pv array is shown. The profile from the band H is shown in Fig. 13.}
\end{figure*}

\begin{figure*}
\epsscale{1.0}
\plotone{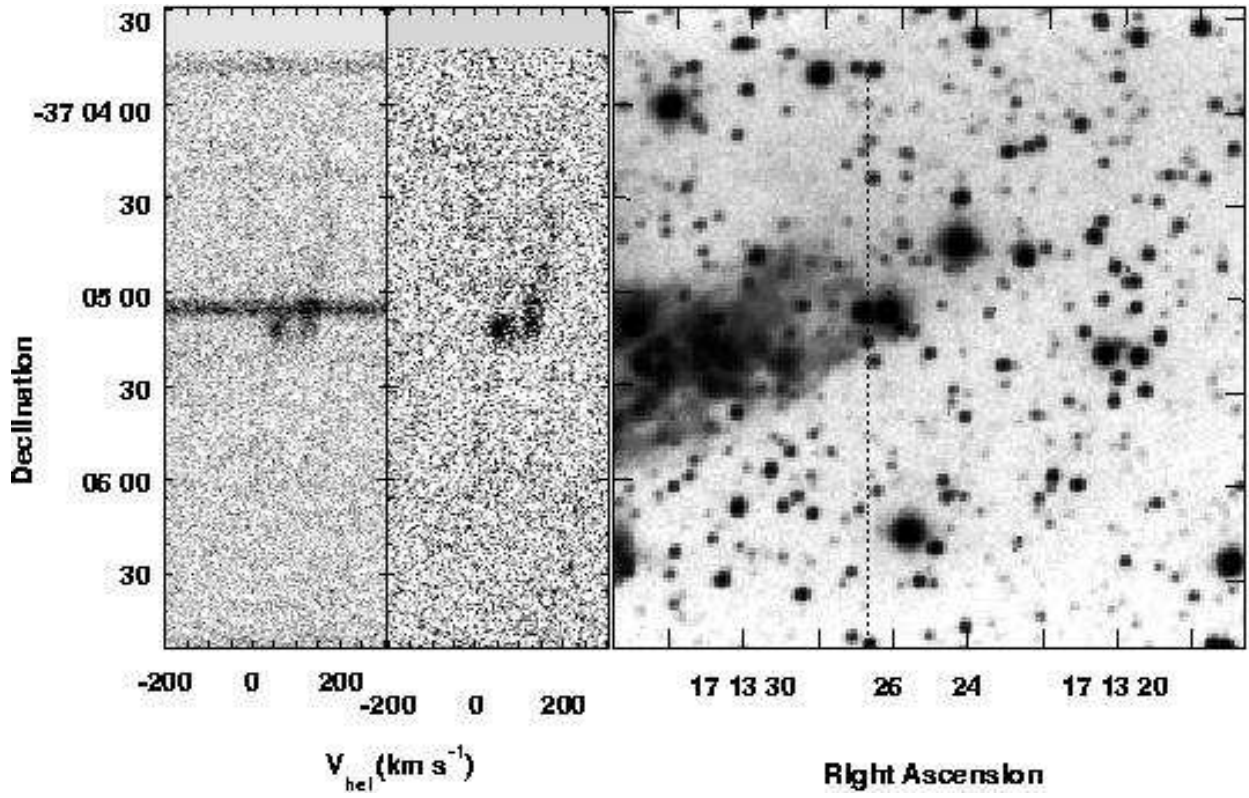}
%\epsfclipon
%\mbox{\epsfxsize=3in\epsfbox[0 0 482 307]{f10.eps}}
\caption{As for Fig. 8 but for slit position 7. The confusing
stellar spectrum (horizontal band) 
is removed from the righthand panel of the pv array
in this display.}
\end{figure*}

\begin{figure*}
\epsscale{0.7}
\plotone{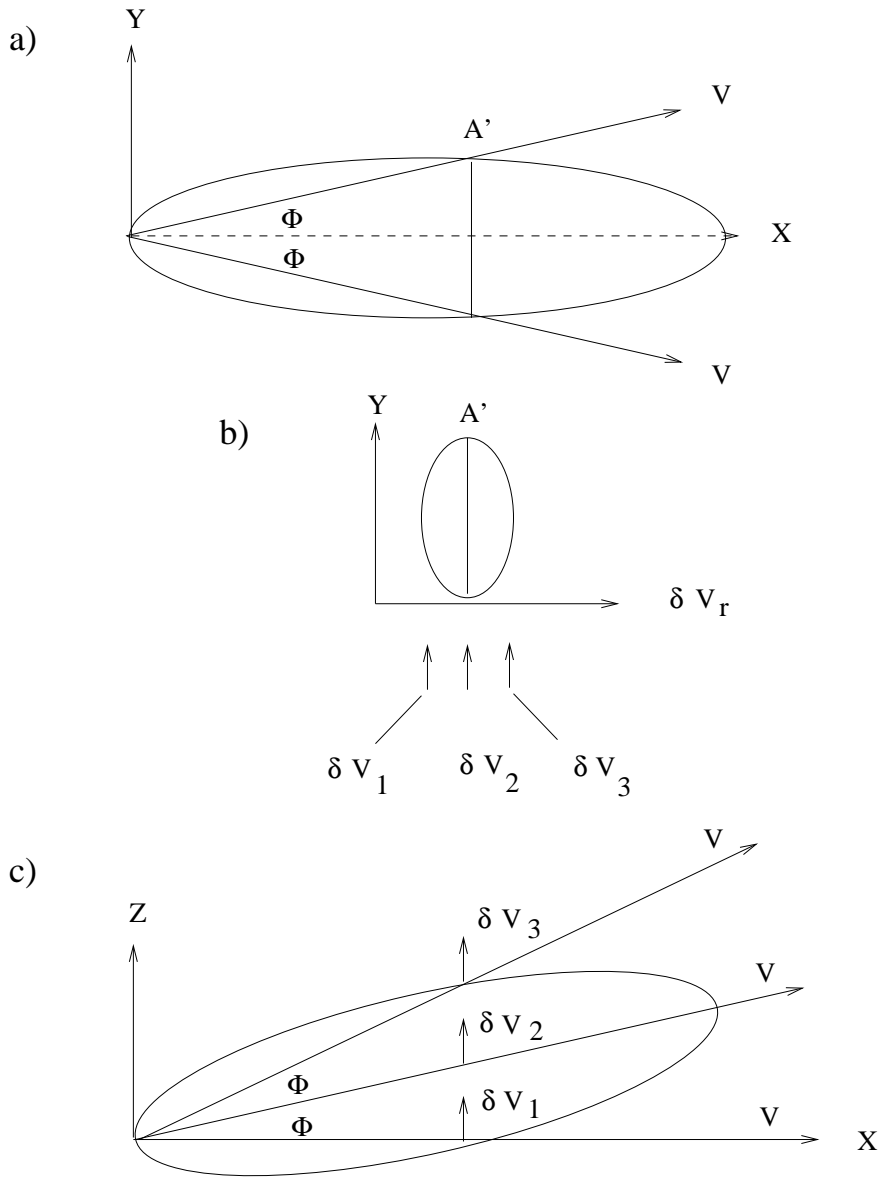}
%\epsfclipon
%\mbox{\epsfxsize=3in\epsfbox[0 0 482 307]{f11.eps}}
\caption{A schematic summary of the measurements from Meaburn
\& Walsh (1980b) along the band marked A' in Fig. 2a is shown. 
In a)  the image (XY plane) of the NW lobe is depicted 
which at A' (see Fig. 2a) is expanding
away from the star at velocity V and angle to the lobe axis of $\phi$.
In b) the measured `velocity ellipse' along the lobe diameter
is sketched where the radial velocity differences are with respect to
\vsys. In c) the geometry of the NW lobe
perpendicuarly (Z dimension) to the plane of the sky is shown 
assuming the NW lobe has a circular section and that
its nearside edge is flowing at V along the 
X--axis in the plane of the sky.} 

 \end{figure*}

\begin{figure*}
\epsscale{0.7}
\plotone{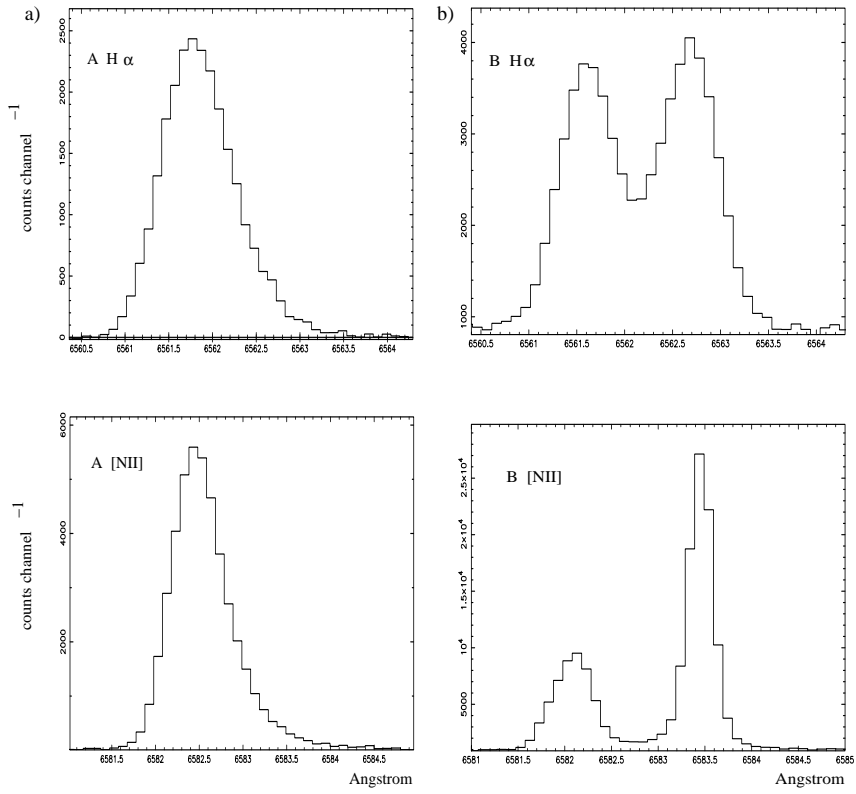}
%\epsfclipon
%\mbox{\epsfxsize=3in\epsfbox[0 0 520 494]{f12.eps}}
\caption{The \ha\ and \niil\ line profiles from the positions
A and B in Fig. 4 are shown in a) and b) respectively.}
 \end{figure*}

\begin{figure*}
\epsscale{0.7}
\plotone{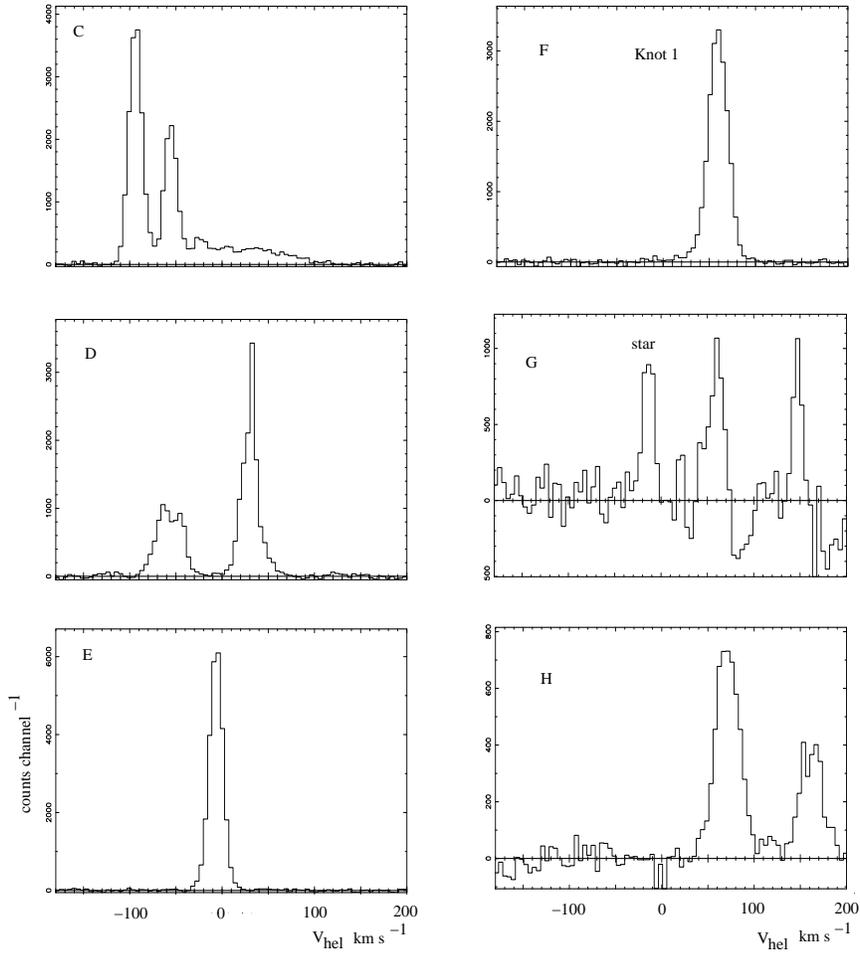}
%\epsfclipon
%\mbox{\epsfxsize=3in\epsfbox[0 0 520 494]{f13.eps}}
\caption{Sample \niil\ line profiles from the positions
C - H in Figs. 5, 6, 7 \& 9 are shown.}
 \end{figure*}

\begin{figure*}
\epsscale{0.9}
\plotone{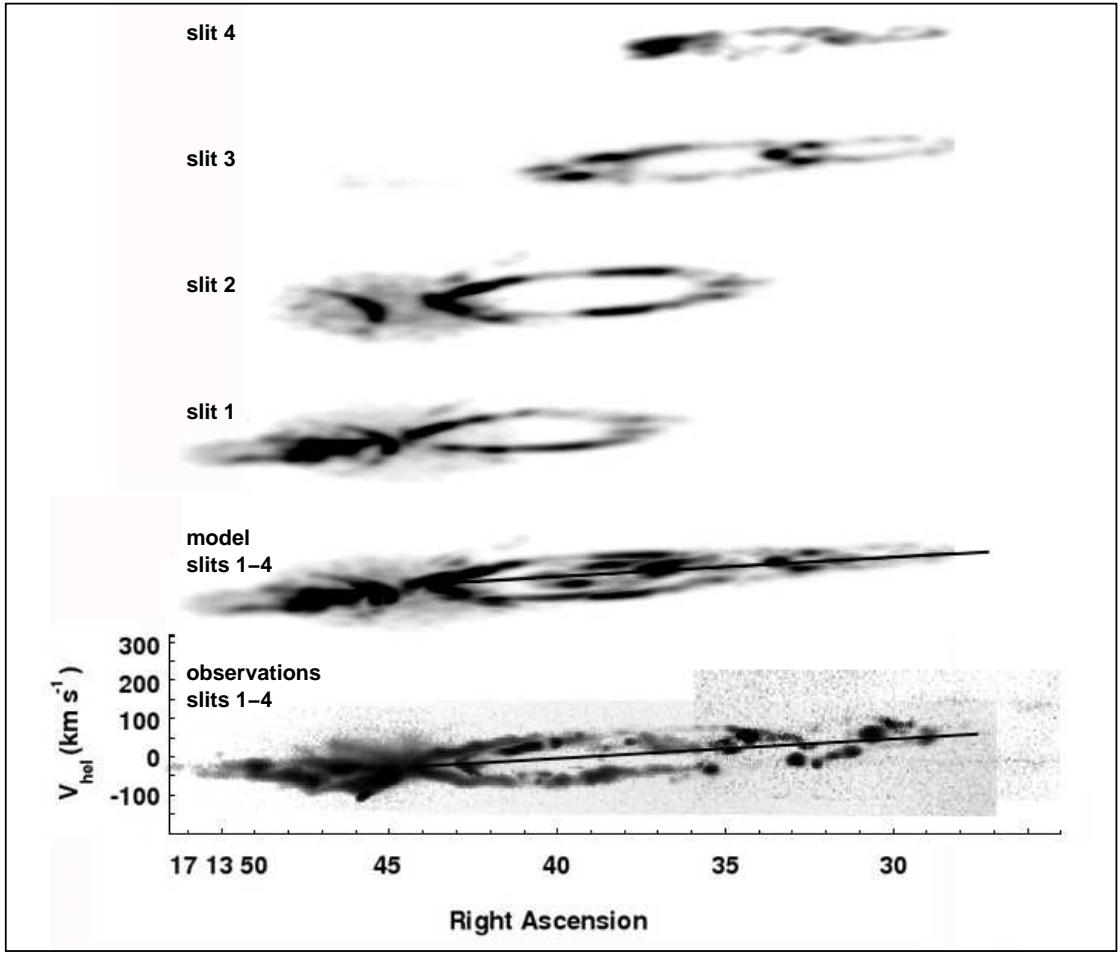}
%\epsfclipon
%\mbox{\epsfxsize=3in\epsfbox[0 0 520 494]{f14.eps}}
\caption{The individual pv arrays predicted by the SHAPE code
for slits 1--4 are shown. Also the superposition of the observed
pv arrays for slits 1--4 is  compared with the model predictions
where the tilted solid line is the expected median if Hubble--type 
flows prevail within the dominant NW lobe. The observed velocities shown
in Fig. 7 of the
faint extremities (far right of the observations for
slits 1--4) can be seen to deviate from this
behaviour significantly.}
 \end{figure*}

\begin{figure*}
\epsscale{0.6}
\plotone{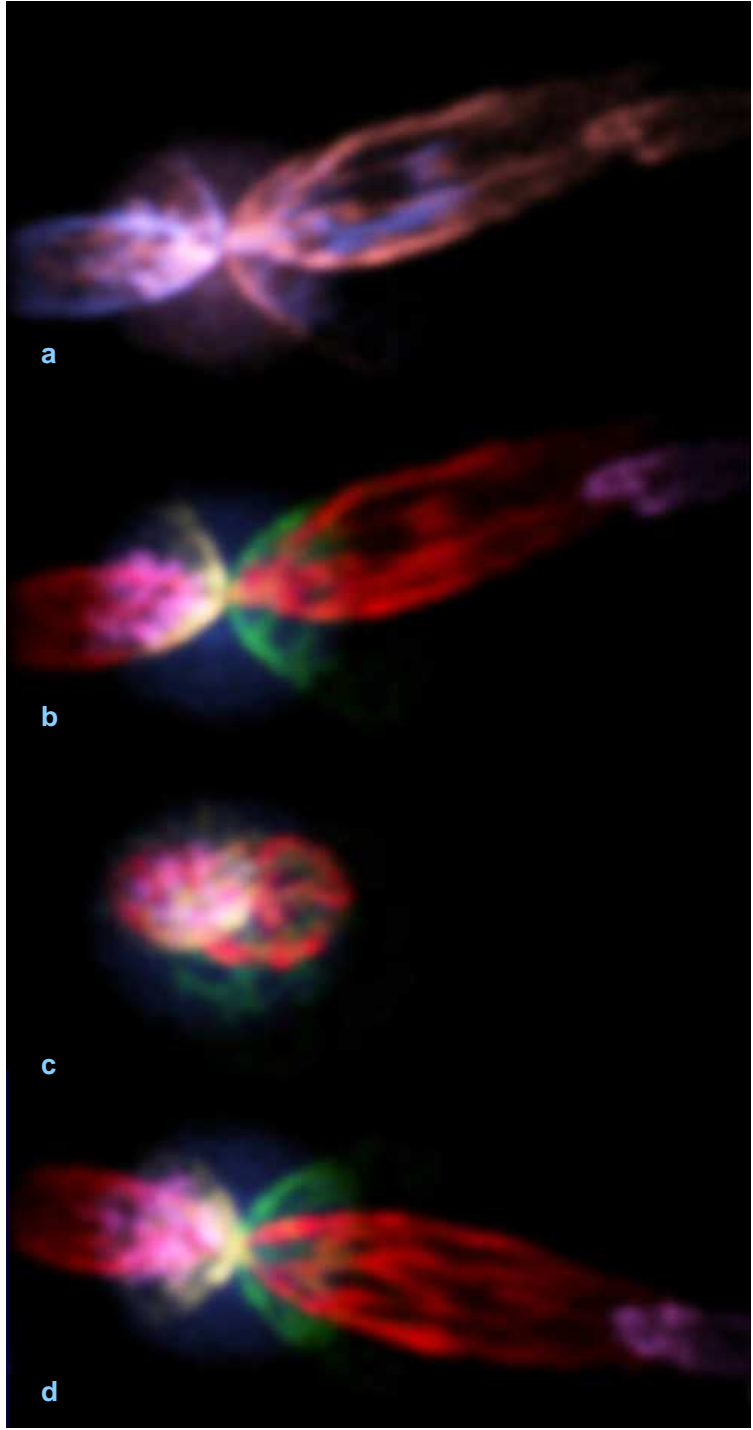}
%\epsfclipon
%\mbox{\epsfxsize=3in\epsfbox[0 0 520 494]{f15.eps}}
\caption{ Images of NGC~6302 predicted by the SHAPE code but viewed from
various directions are shown. a and b are the normal view along the
sight line but where a is color--coded blue/red to indicate the Doppler
shifts and b color--coded arbitarily (as are c and d) to emphasise
separate structures. In c is a NS view in the plane of the sky
and in d an EW view but again in the plane of the sky (as though viewed
from below the image in a).}
 \end{figure*}

\clearpage

\section*{Acknowledgements}

The authors wish to thank the staff of
SPM telescope (Mexico), for their help
during these  observations.
JAL and WS gratefully acknowledge financial support from CONACYT (M\'ex)
grants 32214-E and 37214 and DGAPA-UNAM IN114199 and IN111803. 
MFG is grateful
to PPARC for his research studentship.

\end{document}